\documentstyle[prb,aps,epsfig,floats]{revtex}
\begin{document}
\draft
\wideabs{
\title{Numerical calculation of the combinatorial entropy of
       partially ordered ice}

\author{Bernd A. Berg$^{\rm\,a,b,c,[1]}$ and Wei Yang$^{\rm\,a,d,e}$}

\address{ 
$^{\rm \,a)}$ School of Computational Science, Florida State 
  University, Tallahassee, FL 32306-4120, USA\\
$^{\rm \,b)}$ Department of Physics, Florida State University,
  Tallahassee, FL 32306-4350, USA\\
$^{\rm \,c)}$ John von Neumann-Institut f\"ur Computing,
  Forschungszentrum J\"ulich, 52425 J\"ulich, Germany \\
$^{\rm \,d)}$ Department for Chemistry and Biochemistry, Florida State
  University, Tallahassee, FL 32306-4390, USA\\
$^{\rm \,e)}$ Institute for Molecular Biophysics, Florida State
  University, Tallahassee, FL 32306-4380, USA\\
} 

\date{August 16, 2007} 

\maketitle
\begin{abstract}
Using a one-parameter case as an example, we demonstrate that 
multicanonical simulations allow for accurate estimates of the 
residual combinatorial entropy of partially ordered ice. For the 
considered case corrections to an (approximate) analytical formula 
are found to be small, never exceeding 0.5\%. The method allows one 
as well to calculate combinatorial entropies for many other systems.
\end{abstract}
}
\narrowtext

\section{Introduction}

After the discovery of the hydrogen bond it was recognized that the 
unusual properties of water and ice owe their existence to a combination 
of strong directional polar interactions and a network of specifically
arranged hydrogen bonds \cite{BeFo33,EiKa69,PeWh99}. By experimental 
discovery \cite{Gi33} it was found that ice~I (ordinary ice) has 
in the zero temperature limit \cite{T0limit} a residual entropy 
$S=k\,\ln(W_1)>0$ where $W_1$ is the number of configurations per 
molecule. Subsequently Linus Pauling \cite{Pa35} based the estimate 
$W_1^{\rm Pauling}=3/2$ on the ice rules:
\begin{enumerate}
\item There is one hydrogen atom on each bond (then called hydrogen 
      bond). 
\item There are two hydrogen atoms near each oxygen atom (these three 
      atoms constitute a water molecule). 
\end{enumerate}
Pauling's combinatorial estimate turned out to be in excellent agreement 
with subsequent refined experimental measurements~\cite{Gi36}. This may
be a reason, why it took 25 years until Onsager and Dupuis \cite{OnDu60} 
pointed out that $W_1=1.5$ is only a lower bound, because Pauling's 
arguments for disordered ice omits correlations induced by closed 
loops which are encountered when one requires fulfillment of the ice 
rules for all molecules. Subsequently Nagle \cite{Na65} used a series 
expansion method to derive the estimate $W_1^{\rm Nagle}=1.50685\,(15)$, 
where the error bar is not statistical but reflects higher order 
corrections of the expansion, which are not rigorously under control.

Groundstate entropy calculations by means of multicanonical (MUCA) 
\cite{Be92a} Markov chain Monte Carlo (MCMC) simulations were pioneered 
by Berg and Celik \cite{Be92b}. In a recent paper \cite{Be07} it was 
shown that this approach allows rather easily for an accurate 
finite-size scaling estimate of the residual entropy of 
ice~I, $W_1^{\rm MUCA}=1.50738\,(16)$, where the error bar is 
now purely statistical. In view of eventual higher order finite 
size corrections, which are not included in the MUCA error bar, 
there is satisfactory agreement with Nagle~\cite{Na65}.

With the advent of neutron scattering technology, it became possible 
to measure the actual hydrogen arrangements. Besides fully ordered
and disordered ice phases, there is also evidence for partially 
ordered ice \cite{La73,Lo93,Lo00}. Based on theoretical groundwork
laid by Takagi \cite{Ta48} and Minagawa \cite{Mi81}, an extension of
Pauling's results to partially ordered ice was derived bu Howe and 
Whitworth \cite{Ho87} and greatly generalized by MacDowell et 
al.~\cite{Ma04}. Comparisons with neutron scattering results are 
also made in Ref.~\cite{Ma04}. Besides, the combinatorial residual 
entropy needs to be taken into account when one considers the phases 
of simple models for water/ice~\cite{Ve05}.

As for disordered ice in Pauling's work, correlations are neglected 
in the analytical estimates \cite{Mi81,Ho87,Ma04} of the residual 
entropy of partially ordered ice. The magnitude of corrections is
largely unknown. For instance, before the paper by Howe and Whitworth 
an erroneous equation was used, which was off by up to more than 
50\% for the entropy per molecule. Nagle's method appears to be too 
complicated for these situations. In this article we generalize the 
MUCA approach of Ref.~\cite{Be07} to include partial order and 
calculate numerical corrections to the formula of Howe and 
Whitworth~\cite{Ho87}. Our method is presented in 
section~\ref{sec_method}, details of our numerical implementation 
are given in section~\ref{sec_implement}, followed by the entropy 
estimates in section~\ref{sec_entropy}. Summary and conclusion 
with an outlook on other applications are given in the final 
section~\ref{sec_summary}.

\section{The Method and Preliminaries} \label{sec_method}

\begin{figure}[-t] \begin{center} 
\epsfig{figure=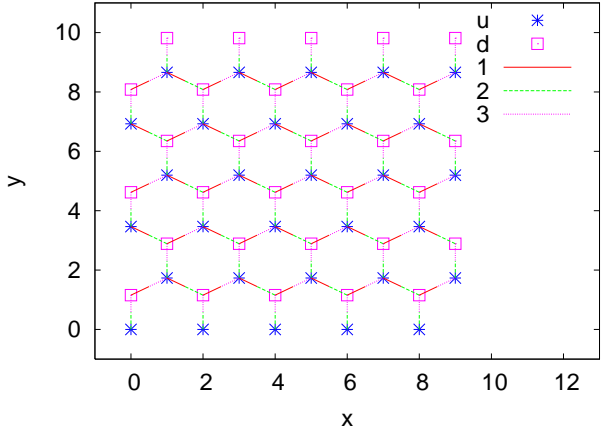,width=\columnwidth} \vspace{-1mm}
\caption{(Color online) Lattice structure of the $z=0$ layer of ice~I. 
The up (u) sites are at $z=1/\sqrt{24}$ and the down (d) sites at 
$z=-1/\sqrt{24}$. For each site three of its four bonds to nearest 
neighbor sites are shown. The fourth bond (to the next layers) is 
in up direction for up and in down direction for down sites. 
\label{fig_icepnt} } 
\end{center} \vspace{-3mm} \end{figure}

As in \cite{Be07} we confine our interest to the hexagonal 
crystal structure of which the $z=0$ layer is is depicted in 
Fig.~\ref{fig_icepnt}. Each oxygen atom is located at the center 
of a tetrahedron and straight lines (bonds) through the sites of 
the tetrahedron point towards four nearest-neighbor oxygen atoms.  
Distances in this figure are given in units of a lattice constant 
$a$ ($a=1$ in the figure(, which is chosen to be the edge length of 
the tetrahedra.  The distance from the center of a tetrahedron to 
one of its sites is $\sqrt{3/8}\,a$ and, hence, the oxygen-oxygen 
distance is $\sqrt{3/2}\,a$. 

This is not the conventional crystallographic definition, but 
convenient for setting up the computer program (see below). For each 
molecule shown one of the surface triangles of its tetrahedron is 
placed in the $xy$-plane. The molecules labeled by u~(up) are then 
at $z=a/\sqrt{24}$ above, and the molecules labeled by d~(down) at 
$z=-a/\sqrt{24}$ below the $xy$-plane, at the centers of their 
tetrahedra. 

\begin{figure} \begin{center}
\psfig{file=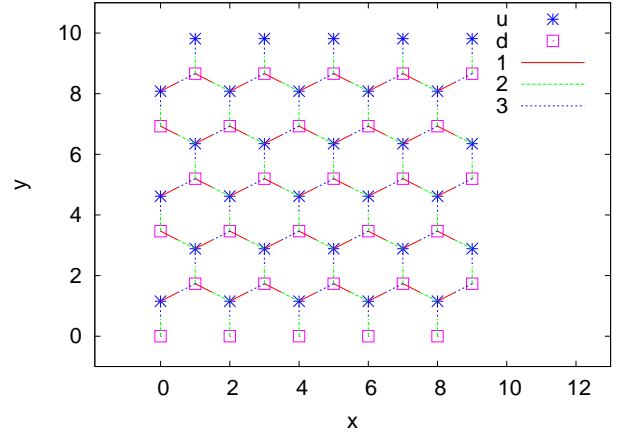,width=\columnwidth} \vspace{-1mm}
\caption{(Color online) Lattice structure of the next, $z=2+2/24$, 
layer of ice~I, above the one of Fig.~\ref{fig_icepnt}.
\label{fig_icepntup} }
\end{center} \end{figure}

We define an ordered reference configuration, which fulfills the ice
rules, by arranging the hydrogen atoms on the bonds in the following 
way:
\begin{enumerate}
\item For $z=i_z 4 a/\sqrt{6}$ and $i_z$ even (as shown in 
      Fig.~\ref{fig_icepnt} for $i_z=0$): For the up oxygens put the 
      hydrogens on bonds 2 and 4, for the down oxygens put them on 
      bonds 1 and~2.
\item For $z=i_z 4 a/\sqrt{6}$ and $i_z$ odd (as shown in 
      Fig.~\ref{fig_icepntup} for $i_z=1$): For the up oxygens put 
      the hydrogens on bonds 3 and 4, for the down oxygens put them 
      on bonds 1 and~3.
\end{enumerate}
This serves as our ordered reference configuration and we denote the 
hydrogen positions in this configuration by $r_b$. 

Following 
\cite{Ho87} we denote the fraction of Hydrogen positions, which
agree with the reference configuration by $f$. The analytical
approximation~\cite{HW} for the residual entropy of configurations 
which fulfill the ice rules is given by
\begin{equation} \label{W1f0}
  W_1^0(f) = \frac{f^{2f} (1-f)^{2(1-f)} 2^{2(f-p)}}
                {p^p (f-p)^{2(f-p)} (1+p-2f)^{(1+p-2f)}}
\end{equation}
with $p=f-2(1-\sqrt{3f^2-3f+1})/3$. The probability that the position 
of a hydrogen atom agrees by chance with the one in the reference
configuration is $1/2$ for disordered ice and Eq.~(\ref{W1f0})
reproduces Pauling's result for this case, $W_1^0(1/2)=1.5$. For 
$f=1$ all hydrogen positions agree with the reference configuration,
and $W_1^0(1)=1$. Correlations due to closed loops of hydrogen bonds
are neglected in the arguments, which lead to Eq.~(\ref{W1f0}). Here
they are included numerically.

The residual entropy of ice~I was calculated in \cite{Be07} by 
performing MUCA simulations for two discrete statistical model, 
which were constructed to satisfy the following properties
[$\beta=1/(kT)$]:
\begin{enumerate}
\item Their total number states (as sampled at $\beta=0$) is known.
\item Generically each model fulfills one of the ice rules, 
      but not the other.
\item In their energy groundstates (reached at large enough $\beta$) 
      each model fulfills both ice rules.
\end{enumerate}
The model, which fulfills ice rule~2 generically is called 6-state
H$_2$O molecule model and has for $N$ molecules a total number of 
$6^N$ states. The model, which fulfills ice rule~1 generically is
called 2-state H-bond model and has $2^{2N}=4^N$ states. Both systems
have similarities with Potts models, so that the lattice labeling
outlined by Fig.~\ref{fig_icepnt} and~\ref{fig_icepntup} allows
one \cite{Be05} to employ simulation methods entirely analogue
to those outlined for Potts models in Ref.~\cite{BBook}. Groundstate
entropy estimates with the 2-state H-bond model turned out to be more
efficient than those with the 6-state H$_2$O molecule model, apparently
because 4 is closer to 1.5 than~6. So we confine our generalization
for partially ordered ice to the 2-state model. To do the same for 
the 6-state model is straightforward, but the simulations are 
expected to be less efficient.

In the 2-state H-bond model \cite{Be07} we allow two positions for each 
hydrogen nucleus on its bond (close toe either one of the two oxygen
atoms, which are connected by the bond). The energy is defined by
\begin{equation} \label{E2}
  E = - \sum_s f_e(s,b^1_s,b^2_s,b^3_s,b^4_s)\,,
\end{equation}
where the sum is over all sites (oxygen atoms) of the lattice and the 
function $f_e$ is given by
\begin{eqnarray} \label{fs}
  &~&f_e(s,b^1_s,b^2_s,b^3_s,b^4_s) = \\ \nonumber &~&
  \cases{2\ {\rm for\ two\ hydrogen\ nuclei\ close\ to}\ s, \cr 
  1\ {\rm for\ one\ or\ three\ hydrogen\ nuclei\ close\ to}\ s,\cr 
  0\ {\rm for\ zero\ or\ four\ hydrogen\ nuclei\ close\ to}\ s. }
\end{eqnarray}
We consider now an additional term 
\begin{equation} \label{Q}
  Q =  \sum_b \delta_{x_b,r_b}\,,
\end{equation}
which is the overlap of the actual positions $x_b$ of the hydrogen 
atoms on the bonds $b$ with the reference positions~$r_b$. The 
canonical ensemble of the extended model is defined by the
Gibbs-Boltzmann weights
\begin{equation} \label{WB}
  \exp(-\beta\,E + h\,Q)\,.
\end{equation}
The coupling parameter $h$ plays pretty much the same role as an
external magnetic field does for the Ising model.

At $\beta=0$ the expectation value of the overlap per link is readily 
computed to be \cite{mBBook}
\begin{equation} \label{Eq}
  \langle q\rangle_{\beta=0} = \langle q\rangle_0 
  = \langle Q\rangle_0 / (2N) = \frac{e^h}{e^h+1}\,,
\end{equation}
and the number of states for which the positions of $K$ hydrogen 
atoms agree with those in the reference configuration is given by 
the binomial factor
\begin{equation} \label{bino}
 B(2N,K) = \left(\matrix{2N\cr K}\right) = \frac{(2N)!}{(2N-K)!\ K!}\ .
\end{equation}
The fraction of correct bonds with respect to the reference 
configuration is given by $f=K/(2N)$. For $K\approx\langle Q\rangle_0$
there will be sufficient statistics so that reweighting of the 
simulation to $\beta =0$ can be used to normalize the spectral 
density via the binomial distribution~(\ref{bino}). For that purpose 
it is convenient to choose $h$ so that $\langle Q\rangle_0$ becomes 
an integer.  Assuming that this is done, we take $K=\langle Q\rangle_0$ 
in the following. 

Using a MUCA weight function 
\begin{equation} \label{WMUCA}
  W_h^{\rm MUCA} = e^{hQ}\,W^{\rm MUCA}(E)
\end{equation}
we can connect the $\beta=0$ region, for which the numbers of states
are known, to the groundstate $E_g=-2N$, for which both ice rules are 
satisfied, and estimate the number of states $n(Q,E_g)$ for $Q$ values 
encountered in the groundstates with sufficient statistics in
$H^{\rm MUCA}(Q,E_g)$ by reweighting:
\begin{equation} \label{MUCA}
  \frac{n(Q,E_g)}{B(2N,K)} =
  \frac{H^{\rm MUCA}(Q,E_g)/w_h^{\rm MUCA}(Q,E_g)}
  {\sum_E H^{\rm MUCA}(K,E)/w_h^{\rm MUCA}(K,E)}\,.
\end{equation}
Here $H^{\rm MUCA}(Q,E_g)$ is the overlap histogram sampled by 
the multicanonical updating in the groundstate ensemble and 
$H^{\rm MUCA}(K,E)$ is the energy histogram sampled for the
fixed value $Q=K$. The reweighting is to $\beta=0$ with $h$ 
unchanged. As the MUCA weights (\ref{WMUCA}) factorize, the
storage requirements are of order $N$ (not $N^2$). 

To obtain a working estimate (see chapter~5.1 of \cite{BBook}) of the 
MUCA weights we use the Wang-Landau recursion \cite{WL01} as explained 
in the next section. The numerical quantities encountered in 
Eq.~(\ref{MUCA}) are often so large that they are not allowed 
by a conventional programing language like Fortran"~77 in Real*8 
precision.  This is overcome by using consistently logarithmic 
coding for which technical detail are explained in \cite{BBook}. 

The actually covered $Q$ range in the groundstate ensemble depends on 
$h$.  Increasing $h$ will shift the range to higher $Q$ values. Doing
so in small steps, and repeating the simulation each time,
\begin{equation} \label{W1MUCA}
  W_1(f) = \frac{1}{N}\,\ln[n(Q,E_g)]
\end{equation}
is obtained for all desired values of $f=Q/(2N)$.

\section{Numerical Implementation} \label{sec_implement}

Using periodic boundary conditions (BCs), our simulations are 
based on the lattice construction of Fig.~\ref{fig_icepnt} 
and~\ref{fig_icepntup}.  Following closely the method outlined 
in chapter~3.1.1 of \cite{BBook} four index pointers from each 
molecule to the array positions of its nearest neighbor molecules 
are constructed along the directions of the bonds as outlined in 
Fig.~\ref{fig_icepnt}. The lattice contains then $N=n_x\,n_y\,n_z$ 
molecules, where $n_x$, $n_y$, and $n_z$ are the numbers of sites
along the $x$, $y$, and $z$ axes, respectively; $i_x=0,\dots,n_x-1$, 
$i_y=0,\dots,n_y-1$, and $i_z=0,\dots,n_z-1$. The periodic BCs 
restrict the allowed values of $n_x$, $n_y$, and $n_z$ to $n_x = 
1,\,2,\,3,\,\dots$, $n_y = 4,\,8, \,12,\,\dots$, and $n_z = 2,\,4,\,6,
\,\dots$~. Otherwise the geometry does not close properly. With the 
inter-site distance $r_{OO}=2.764\,$\AA\ from Ref.~\cite{PeWh99}, 
the physical size of the box is obtained by putting the lattice 
constant to $a=2.257\,$\AA , and the physical dimensions of the 
box are calculated to be $B_x=2n_x\,a$, $B_y=(n_y\,\sqrt{3}/2)\,a$, 
$B_z=(n_z\,4/\sqrt{6})\,a$. In our choices of $n_x$, $n_y$, and 
$n_z$ values we aim within reasonable limitations at symmetrically 
sized boxes.

\begin{table}[tb]
\caption{ Simulation statistics overview. \label{tab_stat}} 
\centering
\begin{tabular}{|c|c|c|c|c|c|} 
 $N$&$n_x$&$n_y$&$n_z$& Statistics & Additional $h_0$-values used \\  
\hline
 128& 4&  8&  4 &$32\times 10^6$ & 0.70 \\ 
\hline
 360& 5& 12&  6 &$32\times 10^6$&    0.65, 0.66, 0.67, 070\\
\hline
 576& 6& 12&  8 &$32\times 3\,10^6$& 0.65, 0.66, 0.67, 070\\
\hline
 896& 7& 16&  8 &$32\times 9\,10^6$& 0.65, 0.66, 0.67, 0.68, 070\\
\hline
1600& 8& 20& 10 &$32\times 32\,10^6$& 0.66, 0.67, 0.68\\
\end{tabular} \end{table} 

The $h$ values at which the simulations are performed are determined 
from initially proposed $h_0$ values in the following way: From 
(\ref{Eq}) we calculate $\langle Q\rangle_0(h_0)$ and determine the 
closest integer $K$. Then the relation (\ref{Eq}) is inverted to
find the value $h$ for which the relation $K=\langle Q\rangle_0(h)$ 
holds. All our simulations use $h_0=0.1$, 0.2, 0.3, 0.4, 0.5, and 0.6. 
Additional $h_0$-values are listed in Table~\ref{tab_stat}, which
gives an overview of our lattices and MUCA production statistics.
The statistics is in sweeps (i.e., updates per molecule) and repeated
32 times. For each of the 32 bins histograms are recorded. To calculate 
error bars they are transformed into jackknife bins along the lines of 
chapter~5.1 of \cite{BBook}.

All calculations can be done by running one 2~GHz PC for about four 
weeks. As the runs at different parameter values are independent,
the real time is considerably shorter when several PCs are available. 
The Wang-Landau recursion \cite{WL01} consumed never more then a few 
percent of a run. Cycling \cite{BBook} and a flatness of 
$H_{\min}/H_{\max}>0.5$ was considered sufficient for iterating 
the Wang-Landau refinement factor. Such a crude flatness is sufficient 
when one does not intend to converge into a reliable estimate of the 
spectral density, as originally proposed in \cite{WL01}, but aims
only at obtaining a working estimate of the MUCA weights. To use the 
Wang-Landau algorithm in this ways as a recursion for the first part 
of a MUCA simulation was suggested in Ref.~\cite{BB03}. 

\begin{figure}[-t] \begin{center} 
\epsfig{figure=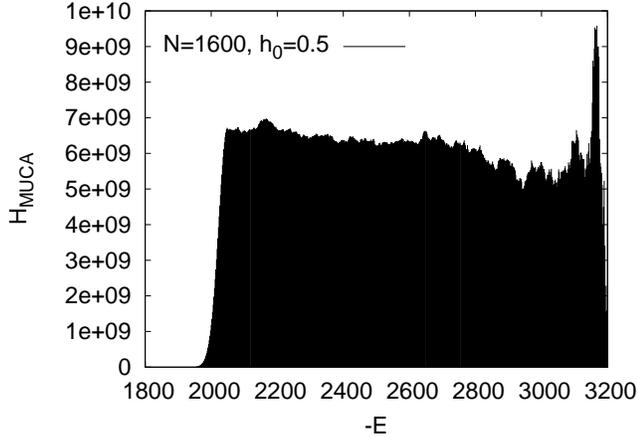,width=\columnwidth} \vspace{-1mm}
\caption{A MUCA energy histogram. \label{fig_50ha1600} }
\end{center} \vspace{-3mm} \end{figure}

\begin{figure}[-t] \begin{center} 
\epsfig{figure=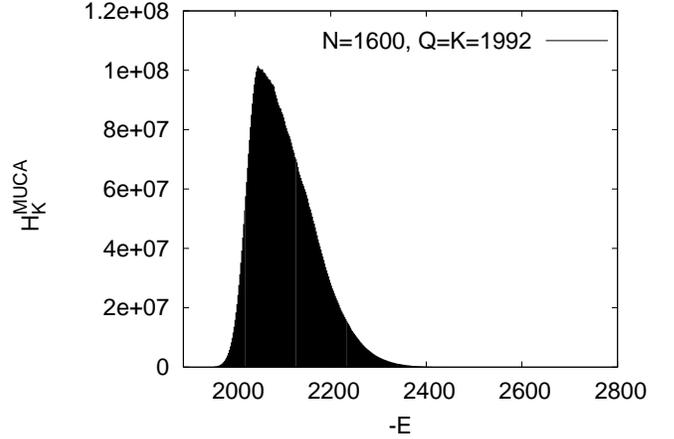,width=\columnwidth} \vspace{-1mm}
\caption{Energy histogram for fixed overlap $Q$.\label{fig_50ham1600}}
\end{center} \vspace{-3mm} \end{figure}

For the $h_0=0.5$ run on our largest lattice the MUCA energy histogram 
of the production part is shown in Fig.~\ref{fig_50ha1600}. The value 
$h_0=0.5$ converts for this lattice to $h=0.500173$ so that $\langle 
Q\rangle_0 = 1992 = K$. In Fig.~\ref{fig_50ham1600} the energy 
histogram is restricted to entries for which $Q=K=1992$ holds. 
It is this histogram, which is reweighted to $\beta=0$ and then 
normalized, so that its sum over energies, the denominator of the 
right-hand side of (\ref{MUCA}), matches the binomial coefficient
~(\ref{bino}). To monitor the entire $(Q,E)$ distribution a histogram 
array $H^{\rm MUCA}(Q,E)$ of size $N^2$ would be needed. In our 
simulations we avoided arrays of size $N^2$ by focusing reweighting 
on one selected $Q$ value, such that an array of size $N$ is sufficient. 
However, this restriction to a microcanonical state was possibly not 
a wise decision. Compared to the analysis of \cite{Be07} 
we find spurious fluctuations and increased error bars. Likely that
could be smoothed out when the full array is available, which would
for our largest lattice still fit into the memory of a PC, and in the
analysis allow to sum over $Q$ for the normalization. As this would
require to repeat all simulations, we cannot pursue this issue
further at this point.

\begin{figure}[-t] \begin{center} 
\epsfig{figure=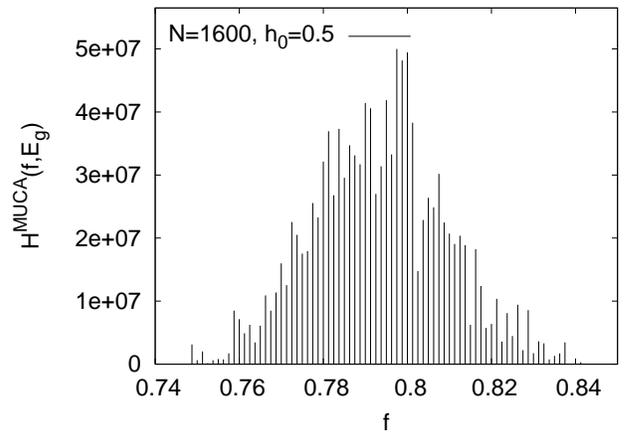,width=\columnwidth} \vspace{-1mm}
\caption{Overlap histogram from groundstates sampled. 
\label{fig_50hma1600} }
\end{center} \vspace{-3mm} \end{figure}

The overlap histogram as measured in the corresponding groundstate 
distribution ($E_g=-2N=-3200$ for this lattice) is depicted in 
Fig.~\ref{fig_50hma1600}. Properly normalized the number of 
configuration per molecule follows from this histogram by using 
Eq.~(\ref{MUCA}) for $f$ values which are sampled with sufficient 
statistics. The cut-off values for sufficient statistics for $f$
were determined from one half of the maximum value of 
$H_{\max}^{\rm MUCA}= \max_f[H^{\rm MUCA}(f,E_g)]$ in the
following way:
\begin{equation} \label{frange}
  f_1\le f \le f_2
\end{equation}
with
\begin{eqnarray} 
  f_1 & = &
  \min_f\left[f;H^{\rm MUCA}(f,E_g)\ge H_{\max}^{\rm MUCA}/2\right]\,,
  \\ f_2 & = &
  \max_f\left[f;H^{\rm MUCA}(f,E_g)\ge H_{\max}^{\rm MUCA}/2\right]\,.
\end{eqnarray}

\section{Entropy Estimates} \label{sec_entropy}

\begin{figure}[-t] \begin{center} 
\epsfig{figure=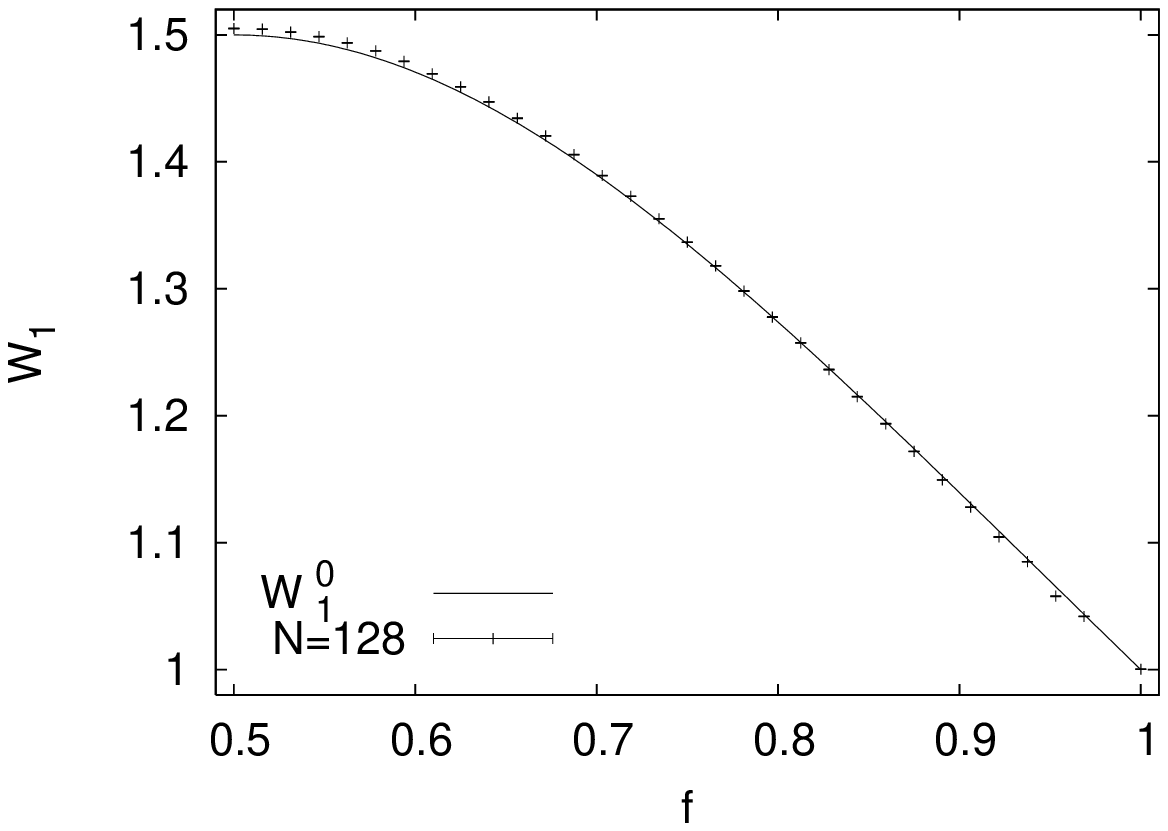,width=\columnwidth} \vspace{-1mm}
\caption{$W_1(f)$ in the approximation (\ref{W1f0}) versus MUCA. 
\label{fig_w1f0128} }
\end{center} \vspace{-3mm} \end{figure}

\begin{figure}[-t] \begin{center} 
\epsfig{figure=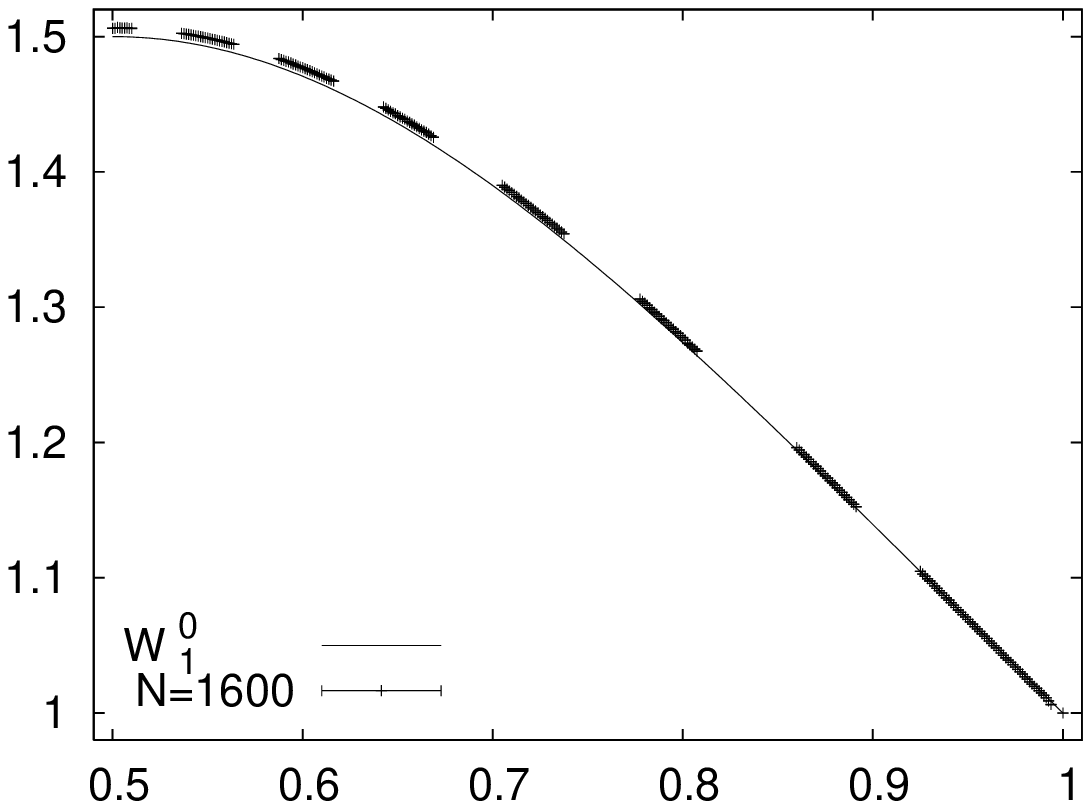,width=\columnwidth} \vspace{-1mm}
\caption{$W_1(f)$ in the approximation (\ref{W1f0}) versus MUCA. 
\label{fig_w1f1600} }
\end{center} \vspace{-3mm} \end{figure}

\begin{figure}[-t] \begin{center} 
\epsfig{figure=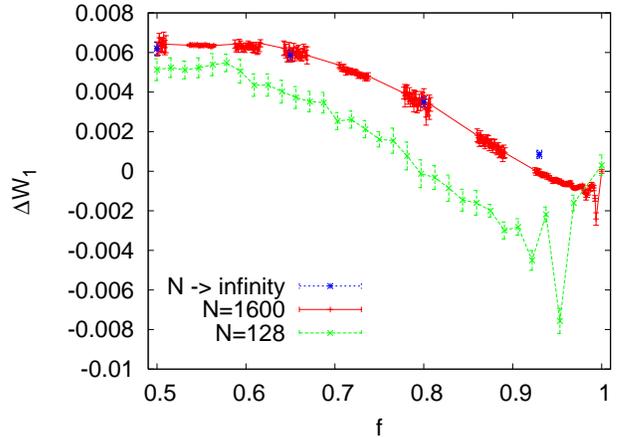,width=\columnwidth} \vspace{-1mm}
\caption{(Color online) Difference between MUCA estimates and the 
approximation (\ref{W1f0}). The lines are only drawn to guide the 
eyes.  \label{fig_w1c} } \end{center} \vspace{-3mm} \end{figure}

Fig.~\ref{fig_w1f0128} compares the approximation $W_1^0(f)$ of
Eq.~(\ref{W1f0}) with the estimates from our smallest lattice
and Fig.~\ref{fig_w1f1600} with the estimates from our largest
lattice. The differences between the numerical results and 
the analytical approximation are in both cases small, but well
outside the range of the numerical error bars. The latter point 
is demonstrated in Fig.~\ref{fig_w1c}, where we plot 
\begin{equation} \label{DW1}
  {\Delta} W_1(f) = W_1^{\rm MUCA}(f;N) - W_1^0(f)
\end{equation}
for $N=128$ and~1600. A feature of Figs.~\ref{fig_w1f1600} 
and~\ref{fig_w1c} is that only patches of $f$ are covered by the 
$N=1600$ data. Each $h_0$ value defines such a patch by means of 
Eq.~(\ref{frange}). The one corresponding to $h_0=0.5$ can be read 
off from Fig.~\ref{fig_50hma1600}: $0.7775 \le f\le 0.8075$. By adding 
simulations for further $h_0$ values the uncovered $f$ regions can be 
filled. We abstained from doing this, because it is only of academic 
interest. Our corrections to the analytical approximation (\ref{W1f0}) 
show that this approximation is sufficiently accurate for practical 
applications, because error bars of experimental entropy estimates 
(e.g., \cite{Ha74}) are much larger than the correction to~(\ref{W1f0}).

\begin{table}[tb]
\caption{ Infinite volume extrapolations of $W_1(f)$ and $\Delta W_1(f)$ 
(the error bars of both quantities are the same). \label{tab_w1}} 
\centering
\begin{tabular}{|c|c|c||c|c|c|} 
$f$ &$W_1(f)$    &$\Delta W_1(f)$&
$f$ &$W_1(f)$   &$\Delta W_1(f)$\\ \hline
0.50&1.50620 (32)&0.00620&0.80&1.27729 (26)&0.00350\\ \hline
0.65&1.44166 (26)&0.00587&0.93&1.09849 (20)&0.00085\\
\end{tabular} \end{table} 

Fig.~\ref{fig_w1c} shows also the finite size corrections to 
$W_1^{\rm MUCA}(f,N)$ encountered when moving from $N=128$ to 
$N=1600$ molecules. These estimates together with those from the 
$N=360$, 576 and 896 lattices allow one to perform infinite volume 
extrapolations $W_1(f) = \lim_{N\to\infty} W_1^{\rm MUCA}(f,N)$. As 
in Ref.~\cite{Be07} for the case $f=0.5$ we fit to the form
\begin{equation} \label{W1fits}
  W_1^{\rm MUCA}(f,N) = W_1(f) + a\,N^{-\theta}\ .
\end{equation}
With the present data the 3-parameter fits turn out to be unstable
and we reduce them to stable 2-parameter fits by using $\theta=0.92$
from \cite{Be07} on input. For four $f$ values the thus obtained
infinite volume extrapolations $W_1(f)$ are collected in 
table~\ref{tab_w1} (error bars are given in parenthesis).

\begin{figure}[-t] \begin{center} 
\epsfig{figure=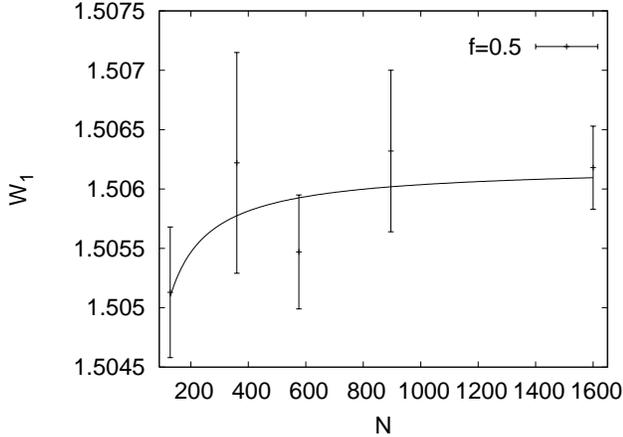,width=\columnwidth} \vspace{-1mm}
\caption{Finite size fit for $W_1(N)$. \label{fig_w1f50} }
\end{center} \vspace{-3mm} \end{figure}

For $f=0.5$ the fit is shown in Fig.~\ref{fig_w1f50}. For the other
$f$ values the shapes of the fits are quite similar. For $f=0.93$
the $N=128$ estimate cannot be included, because it would spoil the
consistency of the fit. Remarkable is that finite size corrections 
for our microcanonically normalized data in Fig.~\ref{fig_w1f50} 
are much smaller than those in the corresponding figure of 
Ref.~\cite{Be07}, where a canonical normalization (summed over
all $Q$ values) of the density of states was used. Further the
sign of the correction is opposite to that Ref.~\cite{Be07}. 
As before, the present estimate is in good agreement with 
Nagle~\cite{Na65}, undershooting now his value slightly, whereas 
the value of \cite{Be07} is overshooting Nagle's estimate somewhat.

\begin{figure}[-t] \begin{center} 
\epsfig{figure=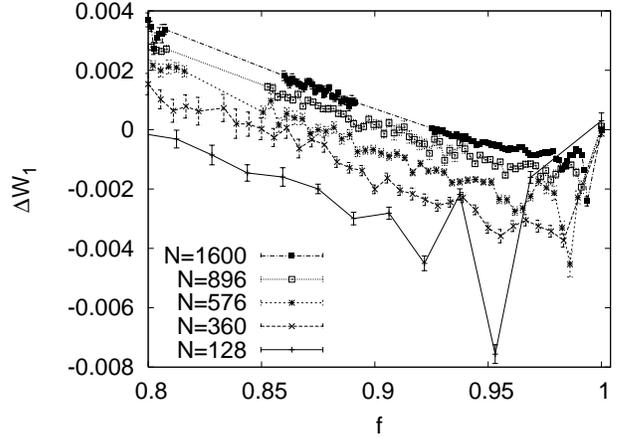,width=\columnwidth} \vspace{-1mm}
\caption{Enlargement of the $f\to 1$ region for $\Delta W_1$.
The lines are only drawn to guide the eyes. \label{fig_w1cto1} } 
\end{center} \vspace{-3mm} \end{figure}

The $W_1(f)$ estimates together with their error bars are also plotted 
in Fig.~\ref{fig_w1c}. Besides for $f=0.93$ they are only visible in 
the color version of this figure, because they fall within the error 
bars of the $N=1600$ data. Interestingly the $f=0.93$ extrapolation
is considerably larger than the $N=1600$ estimate and the sign of the 
correction with respect to the approximation $W_1^0(f)$ (\ref{W1f0})
flipped. While on all our lattices we have for sufficiently large $f$ 
a crossover of the correction from positive to negative, this feature 
may disappear in the $N\to\infty$ limit, so that the corrections are 
ultimately all positive. To illustrate lattice artifacts in the 
$f\to 1$ limit we plot in Fig.~\ref{fig_w1cto1} the $\Delta W_1(f)$ 
values for $f\ge 0.8$. It is clear that the closest values to $f=1$
reflect lattice artifacts and should not be used for the $N\to\infty$ 
approximation. Still estimates for all values of $f$ can be obtained, 
because the $f$ range of the artifacts shrinks $\sim 1/N$.

\section{Summary and Conclusions} \label{sec_summary}

Our main finding is that the corrections to the analytical approximation 
(\ref{W1f0}) are small. As illustrated in table~\ref{tab_w1}, they are
never larger than Nagle's \cite{Na65} already small correction to 
Pauling's \cite{Pa35} value $W_1^{\rm Pauling}=1.5$. For the entropy 
this translates into
\begin{equation} \label{Dentropy}
  \Delta S < \ln\left(W_1^{\rm Nagle}\right) - 
  \ln\left(W_1^{\rm Pauling}\right) \approx 
  \frac{0.00685}{1.5} \approx 0.46\%\ .
\end{equation}
This is beyond the accuracy of nowadays measurements. But who knows
about twenty years ahead? The verification of the correctness of 
predicted correlations beyond the Pauling-like approximation would
be an ultimate confirmation of our understanding of ice.

It is straightforward to include additional parameters in our approach, 
as introduced by the equations of MacDowell et al.~\cite{Ma04}. Each
choice of parameters requires a simulational effort similar to that 
of Ref.~\cite{Be07}. So it would be tedious to map out corrections for 
the entire parameter space. In particular, we did not pursue this 
further, as due to our present results one may conjecture that these 
corrections are also small. If one likes to perform a check for a 
special choice of parameters, for instance because ongoing experimental 
measurements, the details given in our paper should allow researchers
to set up the necessary simulations.

Finally, there may well be applications of our approach to systems
for which corrections to existing approximations are not be small.  
For example, the method allows one to calculate the combinatorial 
entropy of small clusters of hydrogen bonds directly. They are observed 
as formation of ice layers in nanotubes \cite{GRD06} and expected to 
be of importance in the interaction of water with peptides, proteins 
and other biomolecules. Through a better understanding of their entropy 
insights derived from the study of ice may well lead to a better 
understanding of models, which have primarily been constructed to 
reflect interactions of water at room temperature (see \cite{Ve05} 
for an overview).

\acknowledgments
Bernd Berg would like to thank Uli Hansmann and the computational 
biophysics group at the John von Neumann Institut f\"ur Computing
for their kind hospitality during his stay at the Forschungszentrum
J\"ulich.

\clearpage

\begin{thebibliography}{19}

\bibitem{} Correspondence should be addressed to berg@scs.fsu.edu.

\bibitem{BeFo33} J.D. Bernal and R.H. Fowler, J. Chem. Phys. 
                {\bf 1}, 515 (1933).

\bibitem{EiKa69} D. Eisenberg and W. Kauzmann, {\it The Structure and
                 Properties of Water}, Oxford University Press, 
                 Oxford 1969.

\bibitem{PeWh99} V.F. Petrenko and R.W. Whitworth, {\it Physics of
                Ice}, Oxford University Press, Oxford 1999.

\bibitem{Gi33} W.F. Giauque and M. Ashley, Phys. Rev. {\bf 43}, 81 
               (1933).

\bibitem{T0limit} A unique groundstate is expected if one allows for,
                  possibly, astronomically long relaxation times. 
                  Therefore, the residual entropy of ice is not 
                  supposed to violate the third law of thermodynamics.

\bibitem{Pa35} L. Pauling, 
         J. Am. Chem. Soc. {\bf 57}, 2680 (1935). 

\bibitem{Gi36} W.F. Giauque and J.W. Stout, J. Am. Chem. Soc. 
               {\bf 58}, 1144 (1936).

\bibitem{OnDu60} L. Onsager and M. Dupuis, Re. Scu. Int. Fis. `Enrico
                 Fermi' {\bf 10}, 294 (1960).

\bibitem{Na65} J.F. Nagle, 
               J. Math. Phys. {\bf 7}, 1484 (1966). 

\bibitem{Be92a} B.A. Berg and T. Neuhaus, Phys. Rev. Lett. {\bf 68}, 
                9 (1992). 

\bibitem{Be92b} B.A. Berg and T. Celik, Phys. Rev. Lett. {\bf 69}, 
                2292 (1992).

\bibitem{Be07} B.A. Berg, C. Muguruma and Y. Okamoto, Phys. Rev. 
               B {\bf 75}, 092202 (2007).

\bibitem{La73} S.J. La Placa, W.C. Hamilton, B. Kamb, and A. Prakash,
               J. Chem. Phys. {\bf 58}, 567 (1973).

\bibitem{Lo93} J.D. Londono, W.F. Kuhs, and J.L. Finney, J. Chem. 
               Phys. {\bf 98}, 4878 (1993).

\bibitem{Lo00} J.D. Lobban, J.L. Finney, and W.F. Kuhs, J. Chem. 
               Phys. {\bf 112}, 7169 (2000).

\bibitem{Ta48} Y. Takagi, J. Phys. Soc. Jpn. {\bf 3}, 271 (1948).

\bibitem{Mi81} I. Minagawa, J. Phys. Soc. Jpn. {\bf 50}, 3669 (1981).

\bibitem{Ho87} R. Howe and R.W. Whitworth, J. Chem. Phys. 
               {\bf 86}, 6443 (1987).

\bibitem{Ma04} L.G. MacDowell, E. Sanz, C. Vega, and J.L.F. Abascal,
               J. Chem. Phys. {\bf 121}, 10145 (2004).

\bibitem{Ve05} C. Vega, E. Sanz, and J.L.F. Abascal, J. Chem. Phys. 
               {\bf 122}, 114507 (2005).

\bibitem{HW} $W_1=\Omega^{1/N}$ with $\Omega$ and $p$ given by Eq.~(6)
             and (7) of~\cite{Ho87}.

\bibitem{Be05} B.A. Berg, 2005 (unpublished).

\bibitem{BBook} B.A. Berg, {\it Markov Chain Monte Carlo Simulations
                and Their Statistical Analysis}, World Scientific, 
                Singapore, 2004.

\bibitem{mBBook} This is Eq.~(3.70) of \cite{BBook} for $q=2$ and $h=2H$. 
                 Compare also the simulation of chapter 3.3.4.6.

\bibitem{WL01} F. Wang and D.P. Landau, Phys. Rev. Lett {\bf 86}, 
               2050 (2001).

\bibitem{BB03} B.A. Berg, Comp. Phys. Commun. {\bf 153}, 397 (2003).

\bibitem{Ha74} O. Haida, T. Matsuo, H. Suga, and S. Seki, 
               J. Chem. Thermodynamics {\bf 6}, 815 (1974) 

\bibitem{GRD06} N. Giovambattista, P.J. Rossky, and P.G. Debenedetti,
                Phys. Rev. E {\bf 73}, 041604 (2006).

\end{thebibliography}
\end{document}